**On Coherency-Induced Ordering in Substitutional Alloys: I. Analytical**

Dean J. Lee[1] and Jong K. Lee[2+]

[1]Department of Physics, University of Massachusetts, Amherst, MA 01003.
[2]Department of Metallurgical and Materials Engineering, Michigan Technological University, Houghton, MI 49931.

## ABSTRACT

As pointed out by Linus Pauling in his classic work on the relationship between crystal packing and ionic radius ratio, a difference in atomic size can be accommodated more readily by an ordered structure than by a disordered one. Because of mathematical complexity, however, very few works have been reported for substitutional alloys. In this work, coherency-induced ordering in substitutional alloys is examined through a simple model based on a two-dimensional square lattice. Within the assumption of nearest neighbor interactions on a square lattice, both modified Bragg-Williams and Onsager approaches show that coherency strain arising due to atomic mismatch can exert profound effects on order-disorder transitions in substitutional alloys. If the alloy system is elastically homogeneous and Vegard's law is obeyed, the order-disorder transition is of a second-order kinetics. If the atomic mismatches significantly deviate from Vegard's law, however, the transition may become a first-order kinetics, as the configurational free energy surface is composed of double wells. At the transition of a first-order kinetics, the lattice parameter can either increase or decrease upon heating, i.e., the lattice parameter of an ordered state can be less or greater than that of a disordered state. The results of Onsager's approach are independently confirmed with those of the Discrete Atom Method, a Monte Carlo technique predicated upon the combination of statistical mechanics and linear elasticity.

+ To whom all correspondence should be addressed.

## 1. Introduction

It has been well known that one of the driving forces for ordering is the relaxation of elastic strain energy due to difference in atomic size [1-6], as indicated by the change in lattice parameter between ordered and disordered states. As pointed out by Pauling in his classic work on the relationship between crystal packing and ionic radius ratio [3], a difference in atomic size can be accommodated more readily by an ordered structure than by a disordered one. Because of mathematical complexity, however, very few studies have been reported for substitutional alloys [7,8]. In this work, coherency-induced ordering in substitutional alloys is examined through a simple model based on a two-dimensional square lattice. Alloy ordering behavior has usually been examined in terms of chemical bond energies between atomic species as in a quasi-chemical or the Ising model [4-6]. For *interstitial alloys*, Khachaturyan [4] was the first who put forward a strain-induced ordering theory on the basis of elastic interactions between interstitial atoms with a tetragonal misfit strain. Surprisingly however, a couple of strain models are available accounting for strain-influenced ordering in *substitutional alloys*. Vandeworp and Newman [8] considered a Keating-type potential that contains both bond-stretching and bond-bending terms, whereas Beke, Loeff and Bakker [7] applied Eshelby's inclusion theory for the strain energy calculation. These works have brought some light on the role of coherency strain during ordering. However, both models are essentially based on *a priori* assumptions of Vegard's law and homogeneous elasticity, thus their results hardly reveal intricate effects of coherency strain. For example, it is unknown how coherency strain influences an order-disorder transition from a second-order to a first-order kinetics.

As in the Ising model, we make the basic premise that atomic interactions are separable into two parts: *chemical interactions* independent of atomic size and *elastic interactions* solely due to atomic size mismatch. From the viewpoint of a first-principles quantum mechanics approach [9-11], this separation is 'artificial', as the bond strength and bond length depend on the electron density of states, which in turn depends on alloy composition and temperature. Despite its obvious simplicity, the Ising model has been quite successful in providing physical insights on our understanding of phase transitions, thus its basic framework is followed in this work. With advance in high-speed computation, a large number of first-principles investigations are shedding light on the nature of phase equilibria in substitutional alloys. As no theoretical model is perfect, however, extracting coherency-strain effects out of the total enthalpy calculation is not practical in a first-principles approach. It is hoped that a simple analysis such as this work complements quantum mechanics computations or other approaches in our understanding alloy theory.

The article is organized as follows. A modified Bragg-Williams approach [12] is first used to show the effects of coherency strain in ordering of an equi-atomic substitutional alloy with nearest neighbor interactions only. As the Bragg-Williams approach is an over-simplified model, coherency-induced ordering is next examined through Onsager's model which employs a two-dimensional square lattice [13,14]. The results of Onsager's model are then tested with those of the Discrete Atom Method [15,16], a Monte Carlo technique predicated upon the combination of statistical mechanics and linear elasticity. A square lattice with nearest neighbor interactions is mechanically unstable, i.e., its elastic constants are not positive definite. Thus, this work is based on a "hypothetical" crystal, and the work on realistic, stable crystals will be reported in Part II.

## 2. Modified Bragg-Williams Approach

Consider a binary, equi-atomic substitutional AB alloy. The original Bragg-Williams approximation considers chemical bond enthalpies only for the nearest neighbor interactions [12]. The same nearest neighbor interactions are assumed, but A-A bond energy is divided into chemical energy, $E_{AA}$, and strain energy, $S_{AA}$. Defining similar terms for B-B and A-B bonds, the configurational free energy of an $N$-atom system may be written as:

$$G_c = Q_{AA}(E_{AA} + S_{AA}) + Q_{AB}(E_{AB} + S_{AB}) + Q_{BB}(E_{BB} + S_{BB})$$

$$-\frac{Nk_BT}{2}\{2\ln 2 - (1+R)\ln(1+R) - (1-R)\ln(1-R)\} \quad (1)$$

where $Q_{AA}$, $Q_{AB}$, and $Q_{BB}$ are the number of A-A, A-B, and B-B bonds, respectively, and R is the long range order parameter. Typically, R is defined to be the relative fraction of A atoms, which take on the atomic sites of the α sublattice, and is equal to $2f_{A\alpha} - 1$ for an equi-atomic system, where $f_{A\alpha}$ is the fraction of A atoms on α sites. In terms of R, $Q_{AA} = Q_{BB} = Nz(1 - R^2)/8$, and $Q_{AB} = Nz(1 + R^2)/4$, where z is the coordination number equal to 4 for a two-dimensional square lattice.

At an equi-atomic composition, a square lattice displays an elementary form of ordered structure as shown in Fig. 1, where the lattice parameter of a perfect ordered state is designated by 1 + δ. For convenience, let the atoms denoted by open circles be the solvent atoms (A), while the atoms marked with solid circles be the solute atoms (B). The spring constants between A-A, B-B, and A-B interactions are set to $k_1$, $k_2$, and $k_3$, respectively. As the solvent atom is taken to be A, the lattice parameter of pure A at a stress-free state may be set equal to unity. The lattice parameter of pure B is then given by $(1 + \varepsilon_2)$, where $\varepsilon_2$ is the dilatational misfit strain of pure B relative to pure A. Similarly, a misfit strain $\varepsilon_3$ is assigned for the solvent-solute, A-B, bond length at a stress-free state. Note that $\varepsilon_3$ is independent of $\varepsilon_2$ and is not necessarily equal to δ: both $\varepsilon_3$ and $k_3$ should be regarded as physical entities conjugate to $E_{AB}$. In the limit of nearest neighbor interactions only, the strain energy of each bond could be approximated as follows: $S_{AA} = k_1\delta^2/2$, $S_{BB} = k_2(\delta - \varepsilon_2)^2/2$, and $S_{AB} = k_3(\delta - \varepsilon_3)^2/2$. This is based on the assumption that no atoms are allowed to deviate from the lattice sites and thus the structure must maintain the symmetry of a square lattice. However, its lattice parameter is allowed to vary with 1 + δ. If the spring constants are uniform at $k_1 = k_2 = k_3 = k$, it means an elastically homogeneous system. For the purpose of obtaining analytical expressions, but with no loss of main features, homogeneous elasticity is hereafter assumed.

Obviously, a square lattice with only nearest neighbor interactions is mechanically unstable, and any atomic movements away from the sites would transform the lattice to a stable triangular lattice. Nonetheless, we proceed with this hypothetical square lattice, as it will be shown to provide interesting physical insights. When the strain energy terms of $S_{AA}$, $S_{BB}$, and $S_{AB}$ are introduced into Eq. (1) and the configurational free energy is minimized with respect to δ at a given temperature, δ is found to be:

$$\delta = \{\varepsilon_2 + 2\varepsilon_3 + R^2(2\varepsilon_3 - \varepsilon_2)\}/4 \quad (2)$$

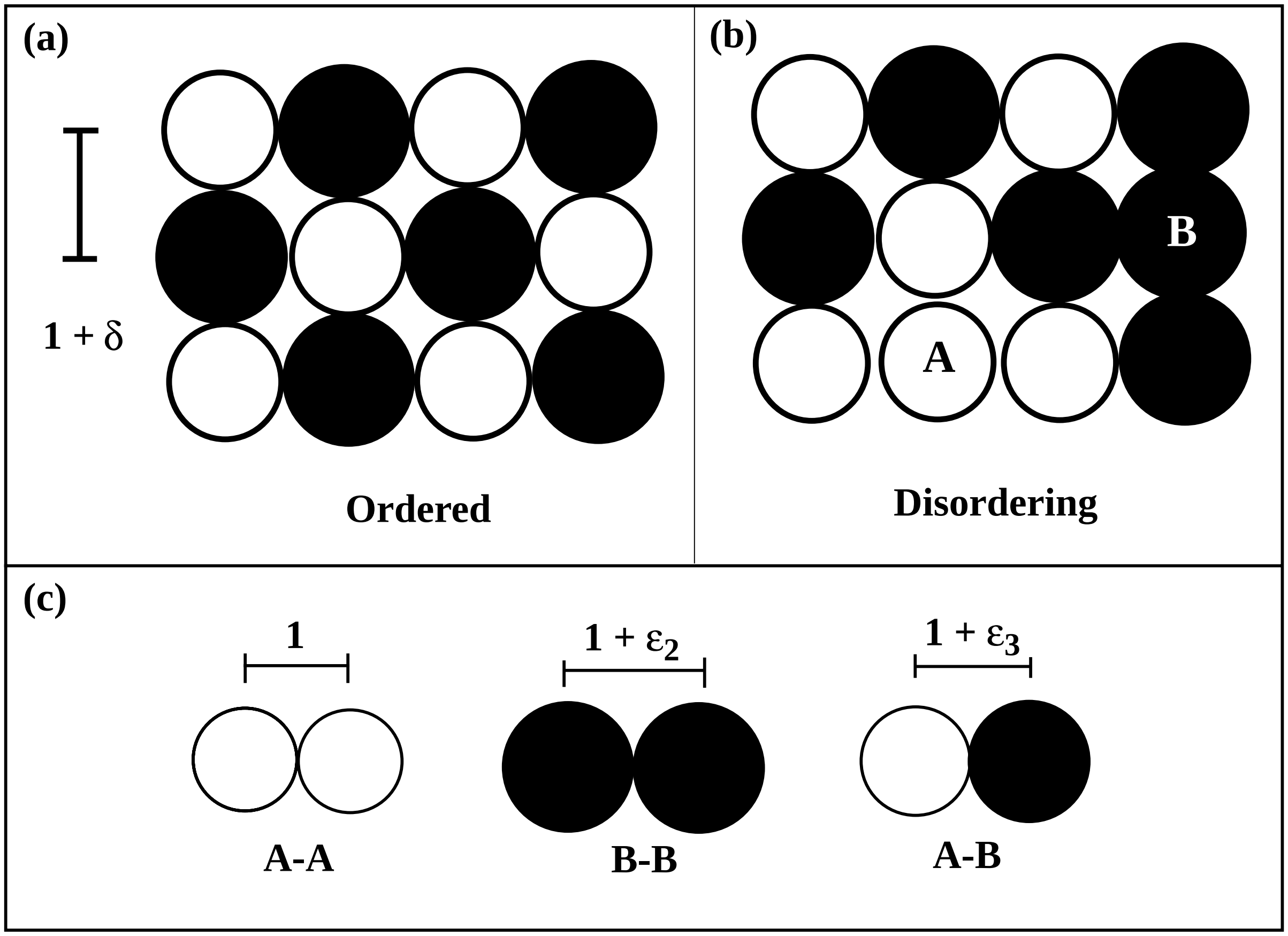

Fig. 1: An elementary form of an ordered structure in a two-dimensional square lattice. (a) a perfect ordered state with a lattice parameter equal to $1 + \delta$. (b) a state of disordering with a lattice parameter different from $1 + \delta$. (c) lattice parameters at stress-free states.

Notice an interesting result that the lattice parameter becomes independent of R when the system follows Vegard's law, that is, $2\varepsilon_3 = \varepsilon_2$. Substituting Eq. (2) into Eq. (1) and introducing a reduced free energy, *g* (in units of $Nzk\varepsilon_2^2/64$), and a reduced temperature, $\theta$ (in units of $zk\varepsilon_2^2/16k_B$):

$$g = -R^2\left(\frac{16\Omega_c}{k\varepsilon_2^2} + 2\right) - R^4\left(\frac{2\varepsilon_3}{\varepsilon_2} - 1\right)^2 + 2\theta\{(1+R)\ln(1+R) + (1-R)\ln(1-R)\} \qquad (3)$$

where $\Omega_c$ is the usual chemical ordering energy equal to $(E_{AA} + E_{BB})/2 - E_{AB}$. The first term with $R^2$ is a familiar one: it is responsible for a second-order kinetics for order-disorder transition in a AB binary alloy. If the system obeys Vegard's law and thus the $R^4$ term vanishes, the ordering energy is simply the sum of both a chemical and a strain origin.

If $2\varepsilon_3 \neq \varepsilon_2$, the $R^4$ term may force the system into a class of first-order kinetics during order-disorder transition, as it induces two energy wells in the free energy surface. In Fig. 2, the long range order parameter, R, is plotted as a function of temperature, $\theta$, for the two different cases of $2\varepsilon_3/\varepsilon_2$

= 1 and $2\varepsilon_3/\varepsilon_2$ = -1. For simplicity, the chemical ordering energy, $\Omega_c$, is set to zero for both cases. The critical temperature for order-disorder transition is marked with $\theta_c$. In (a), the transition is a second-order, whereas (b) demonstrates a first-order kinetics. Besides the discontinuity in R, the first-order kinetics of the case (b) can also be tested with a change in the lattice parameter: the low-temperature, ordered structure has R = 0.97 at $\theta = \theta_c$, and thus $\delta = 0.94\varepsilon_3$ with $2\varepsilon_3/\varepsilon_2 = -1$ from Eq. (2), while the disordered, high-temperature phase has R = 0, which yields $\delta = 0$. Therefore, depending on the sign of $\varepsilon_3$, the lattice parameter may increase or decrease upon disordering at the transition temperature.

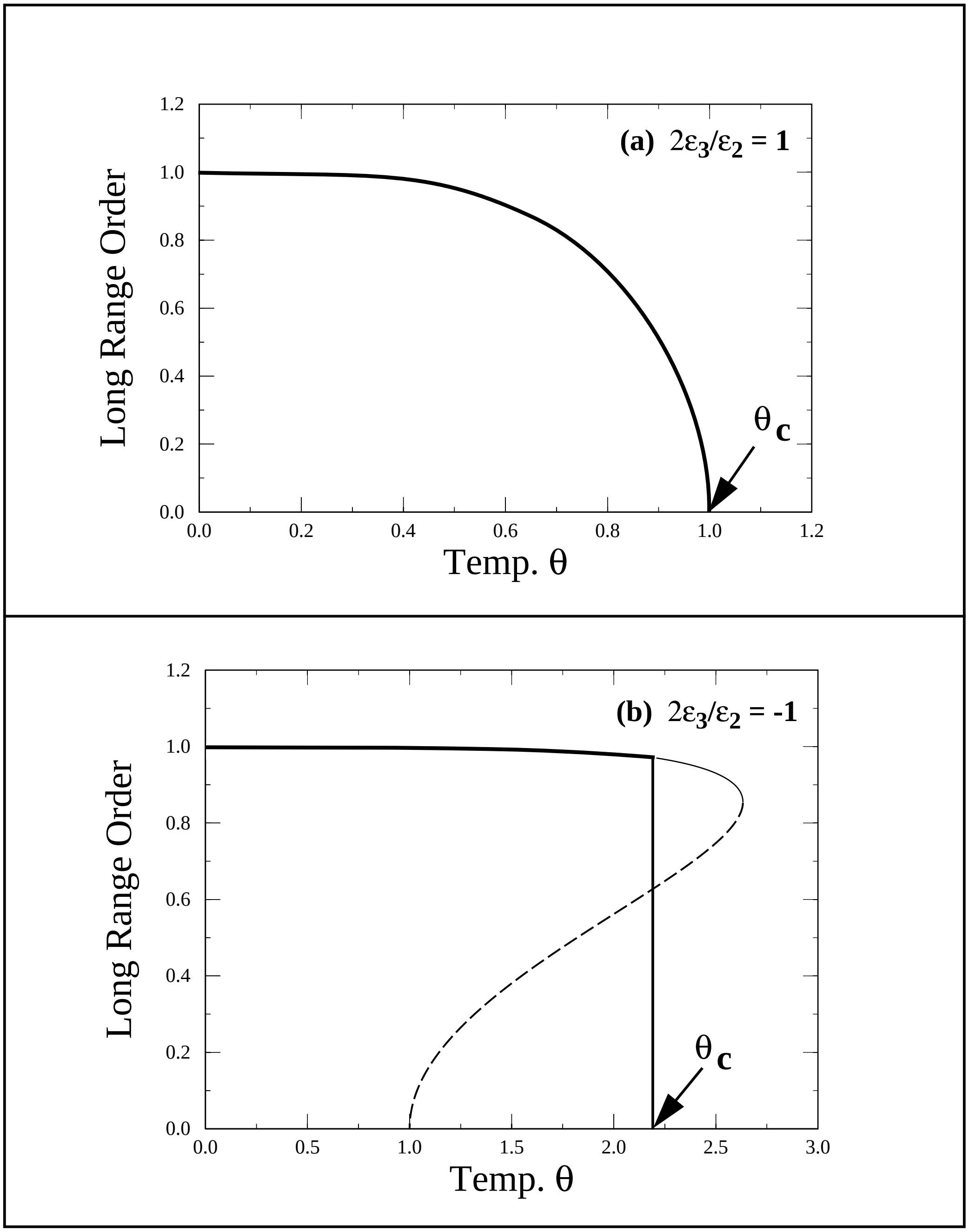

Fig. 2: Long range order parameter, R, vs. reduced temperature, θ, for an equi-atomic AB alloy based on a square lattice with nearest neighbor interactions only. The driving force for ordering is elastic strain energy due to coherency strain. In (a) a system of second-order kinetics obeying Vegard's law with $2\varepsilon_3/\varepsilon_2$ = 1, and (b) a system of first-order kinetics with $2\varepsilon_3/\varepsilon_2$ = -1.

In Fig. 2, (a) represents a continuous bifurcation phenomenon, whereas (b) represents a discontinuous or jump bifurcation phenomenon. Both curves stand for $\partial g/\partial R = 0$ for a given $\theta$. In (b), however, the curve is divided into three parts: the heavy solid curve representing an absolute minimum in g, the thin solid indicating a local minimum, and the dashed one standing for a local maximum. The local minimum free energy on the thin solid curve is greater than that of a complete disordered state, i.e., g = 0 with R = 0 (see Eq. 3). A jump bifurcation phenomenon such as (b) is a ubiquitous behavior observed in many first-order transitions [17]. For example, coherency strain energy influences the activation barrier for nucleation in solid-state phase transformations in a way very similar to this case [18].

For an $A_3B$ alloy (such as $Ni_3Al$), the Bragg-Williams theory predicted a first-order kinetics for the order-disorder transition [12], but its origin is attributed to the *configurational entropy.* Here, coherency strain effects demonstrate a first-order kinetics, but now its origin resides in the *configurational enthalpy.* When Vegard's law is followed, the critical temperature is simply given by the sum of the two individual cases, i.e., the chemical and the strain ordering contributions. If Vegard's law is broken and the kinetics becomes of a first-order, however, Eq. (3) displays that there would be a coupling effect between the chemical and the strain ordering energy, and the transition temperature will be different from the sum of the individual cases. For example, a system with $2\varepsilon_3/\varepsilon_2 = -1$ and $16\Omega_c/k\varepsilon_2^2 = 2$ yields $\theta_c = 2.98$, as compared to the sum of the individual cases, $\theta_c = 1$ from $\Omega_c$ ($= k\varepsilon_2^2/8$) and $\theta_c = 2.2$ from $2\varepsilon_3/\varepsilon_2 = -1$. The Bragg-Williams approach is an over-simplified model and its predicted order-disorder temperature is quite different from the solution of Onsager [5]: for example, the exact transition temperature for the second-order kinetics shown in Fig. 2(a) is $\theta_c = 0.567$. Other exact solutions are examined below.

## 3. Modified Onsager Approach

By expressing chemical interaction energies as $E_{AA} = a + b + c$, $E_{BB} = a - b + c$, and $E_{AB} = -a + c$, Onsager [13,14] derived the configurational free energy per atom for an equi-atomic system based on a two-dimensional square lattice as:

$$G_c = -k_B T\left[\ln 2 + 0.5\ln y + \frac{1}{8\pi^2}\int_0^{2\pi}\int_0^{2\pi}\ln\left\{y + \frac{1}{y} - \cos\varphi - \cos\varsigma\right\}d\varphi d\varsigma\right] + 2c \qquad (4)$$

where y = sinh $(2a/k_B T)$. For an ordering system with $E_{AA} = E_{BB} = 0$ and $E_{AB} = -\Omega_c$, it follows that $a = (E_{AA} + E_{BB} - 2E_{AB})/4 = \Omega_c/2$, $b = (E_{AA} - E_{BB})/2 = 0$, and $c = (E_{AA} + E_{BB} + 2E_{AB})/4 = -\Omega_c/2$. Since Onsager's system represents a grand canonical ensemble, the *equi-atomic condition*, i.e., $N_A = N_B$, assumes that the chemical potentials of A and B are appropriately controlled, and furthermore implies that Eq. (4) is still valid even if $b \neq 0$ as the contribution to the free energy of the system has the product form $2b(N_A - N_B)$.

We now assume the same conditions except that, as in the modified Bragg-Williams approach, each atomic bond is made of both chemical and strain energy. Thus the three bond energies become $k_1\delta^2/2$, $k_2(\delta - \varepsilon_2)^2/2$, and $-\Omega_c + k_3(\delta - \varepsilon_3)^2/2$ for A-A, B-B, and A-B interaction, respectively. For an elastically homogeneous system with a spring constant equal to $k$, expressions for $a$, $b$ and $c$ are:

$$\left.\begin{aligned} a &= \frac{\Omega_c}{2} + \frac{1}{8}k\left\{-2\delta(\varepsilon_2 - 2\varepsilon_3) + \varepsilon_2^2 - 2\varepsilon_3^2\right\} \\ b &= \frac{1}{8}k\left\{4\delta\varepsilon_2 - 2\varepsilon_2^2\right\} \\ c &= -\frac{\Omega_c}{2} + \frac{1}{8}k\left\{-2\delta(\varepsilon_2 + 2\varepsilon_3) + \varepsilon_2^2 + 2\varepsilon_3^2 + 4\delta^2\right\} \end{aligned}\right\} \qquad (5)$$

Onsager's free energy expression, Eq. (4), is then still valid except that $G_c$ is now a function of both $\Omega_c$ and $\delta$ through $a$ and $c$ of Eq. (5). Because of the equi-atomic condition, $b$ exerts no influence, but the free energy must be minimized with respect to the lattice parameter, that is, $\delta$, for a given set of $\Omega_c$, $k$, $\varepsilon_2$, $\varepsilon_3$, and T. Obviously, controlling chemical potentials of A and B such that $N_A = N_B$ could be difficult to achieve with arbitrary $\varepsilon_2$ and $\varepsilon_3$ values, and thus the system should be regarded a hypothetical one as pointed out before.

## 4. Discrete Atom Method

According to classical statistical mechanics, the configurational free energy is equal to $-k_BT \ln Z_q$, where $k_B$ is Boltzmann's constant and $T$ is the absolute temperature. $Z_q$ is the configurational partition function which is given by:

$$Z_q = \int exp(-\Phi/k_BT)d\vec{q} \qquad (6)$$

In Eq. (6), $\vec{q}$ indicates atomic coordinates. For a square lattice with nearest neighbor interactions only, the Hamiltonian, $\Phi$, may be written as:

$$\Phi = \frac{1}{2}\sum_{i=1}^{N}\sum_{j=1}^{4} E_{ij} + \frac{1}{4}\sum_{i=1}^{N}\sum_{j=1}^{4} k_{ij}(r_{ij} - a_{ij})^2 \qquad (7)$$

where $N$ is the total number of atoms, $E_{ij}$ is the chemical bond energy between $i$-th atom and its $j$-th nearest neighbor, $k_{ij}$ is the spring constant, $r_{ij}$ is the distance between $i$-th and $j$-th atom, and $a_{ij}$ is the value of $r_{ij}$ if all the atoms were at their stress-free state. For the present model square lattice of homogeneous elasticity, $E_{ij}$ is equal to either 0, 0, or $-\Omega_c$ depending on A-A, B-B, or A-B bond, respectively, and $k_{ij} = k$. Likewise, $a_{ij}$ is equal to 1, $1 + \varepsilon_2$ or $1 + \varepsilon_3$ depending on A-A, B-B, or A-B bond. Since the symmetry of a square lattice is preserved, $r_{ij}$ is equal to $1 + \delta$. Topological evolution is then examined through a Monte Carlo process, which generates a Boltzmann-weighted chain of configurations for the given system. Thus, a pair of A and B atoms are randomly selected and their site exchange is tested with $\exp(-\Delta\Phi/k_B T)$, where $\Delta\Phi$ is the energy difference between the new and old configurations. In addition to atomic exchanges, the lattice parameter, i.e., $\delta$, must be allowed to change for strain relaxation. During a simulation, the total number of atoms, $N$ (= $2N_A = 2N_B$), is maintained at a fixed number under a periodic boundary condition, thus the system represents essentially a canonical ensemble if small changes in the lattice parameter are ignored.

## 5. Comparison between Onsager Approach and DAM

To check the validity of the Discrete Atom Method (DAM), the Ising model with $E_{AB} = -\Omega_c = -500k_B$ and $E_{AA} = E_{BB} = \varepsilon_2 = \varepsilon_3 = 0$, i.e., with chemical interactions only, is tested for the order-disorder transition temperature in a 512-atom system ($N_A = N_B = 256$). A DAM result finds the critical temperature at 575 K: this is close to Onsager's exact solution of 567 K through Eq. (4). As a further check, a system obeying Vegard's law with $\varepsilon_2 = 2\varepsilon_3 = 0.02$, $k = 1\text{x}10^7 k_B$, and $E_{AA} = E_{BB} = E_{AB} = 0$ is examined through both the modified Onsager Approach (OA) and the DAM. For this case, OA yields the optimum $\delta$ equal to $\varepsilon_3$ and its critical temperature is again at 567 K, as its effective ordering energy is equal to $500k_B$ from $(S_{AA} + S_{BB})/2 - S_{AB}$. In Fig. 3, specific heat, $C_v$, (in units of $Nk_B$) is plotted as a function of temperature in (a), and its derivative, $dC_v/dT$, is displayed in (b). In (a), the dashed curves are the $C_v$ values predicted by OA. In the DAM, specific heats are calculated through the usual statistical ensemble sampling [12], i.e., $C_v k_B T^2 = \langle E^2 \rangle - \langle E \rangle^2$, where $E$ is the configurational energy. As demonstrated in Fig. 3(b), DAM's order-disorder transition point, $T_c$, is determined by the sign change in $dC_v/dT$, which is also evaluated with an ensemble sampling technique.

To minimize the effect of statistical fluctuations associated with a Monte Carlo process, several runs on both heating and cooling cycle are performed near the critical temperature. The open circles mark raw data points, to which solid curves are fitted. Compared to OA's 567 K, the DAM work yields 580 K (marked by $T_c$) as the critical temperature, and thus its specific curve is somewhat shifted to the right compared to the dashed curve of OA. Additionally, the DAM shows no sign of singularity at the critical temperature, which should be expected from an atomistic simulation. For the small discrepancy in the critical temperature between the two methods, the $N$-size effect in the DAM ($N = 512$) is regarded as the main cause among others. In addition to $C_v$, the long range order parameter R (= $2f_{A\alpha} - 1$, as defined before), a short range order parameter $\sigma$ (= $\langle Q_{AB} \rangle / N - 1$), and configurational entropy are also monitored during the DAM simulations. Configurational entropy is approximated as the ensemble average for site configurational entropies. In Fig. 4, the configurational entropies from both OA and DAM are compared for the same system obeying Vegard's law with $\varepsilon_2 = 2\varepsilon_3 = 0.02$. Again, the dashed curve represents OA and the solid curve fitted to open circles stands for DAM. Clearly, near the order-disorder critical temperature, the DAM configurational entropy shows an overestimation compared to that of OA: one reason for this overestimation is that any migration or rotation of an ordered (or clustered) island makes a contribution to the site configurational entropies. Although DAM's entropy shows a steep increase near the critical temperature, no sign of first-order kinetics is detected. For the lattice parameter, DAM provides $\delta = 0.01 \pm 0.0001$ in a good agreement with OA's prediction, $\delta = \varepsilon_3 = 0.01$.

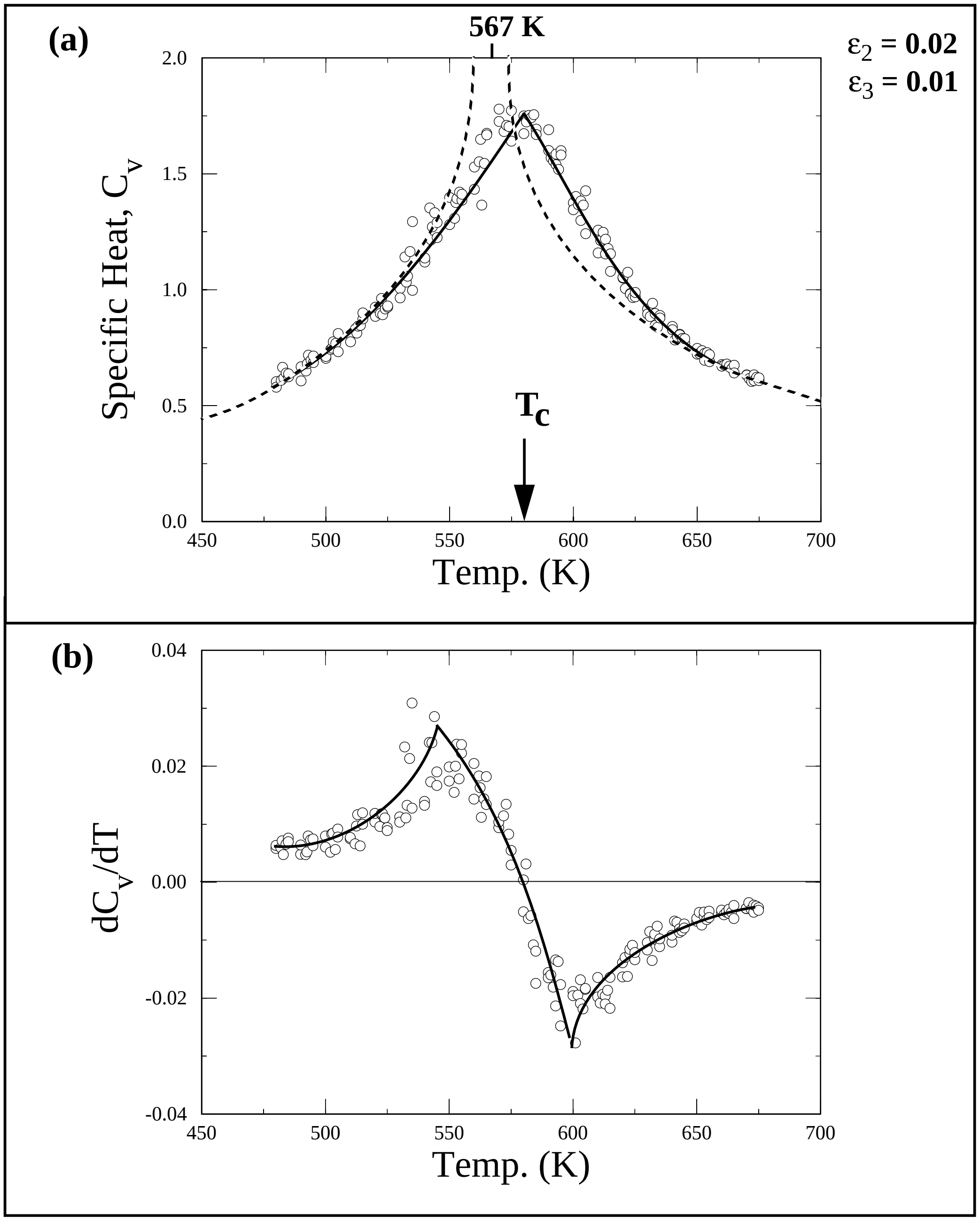

Fig. 3: Order-disorder transition behavior for a two-dimensional square lattice system with $\varepsilon_2 = 2\varepsilon_3 = 0.02$, $k = 1\text{x}10^7 k_B$, and $E_{AA} = E_{BB} = E_{AB} = 0$. In (a), specific heat is plotted as a function of temperature. The dashed curves indicate Onsager's solution, while the solid curve fitted to open circles represents DAM results. $T_c$ marks DAM's transition temperature, 580 K. (b) presents a plot of $dC_v/dT$ versus T, which is used to determine $T_c$.

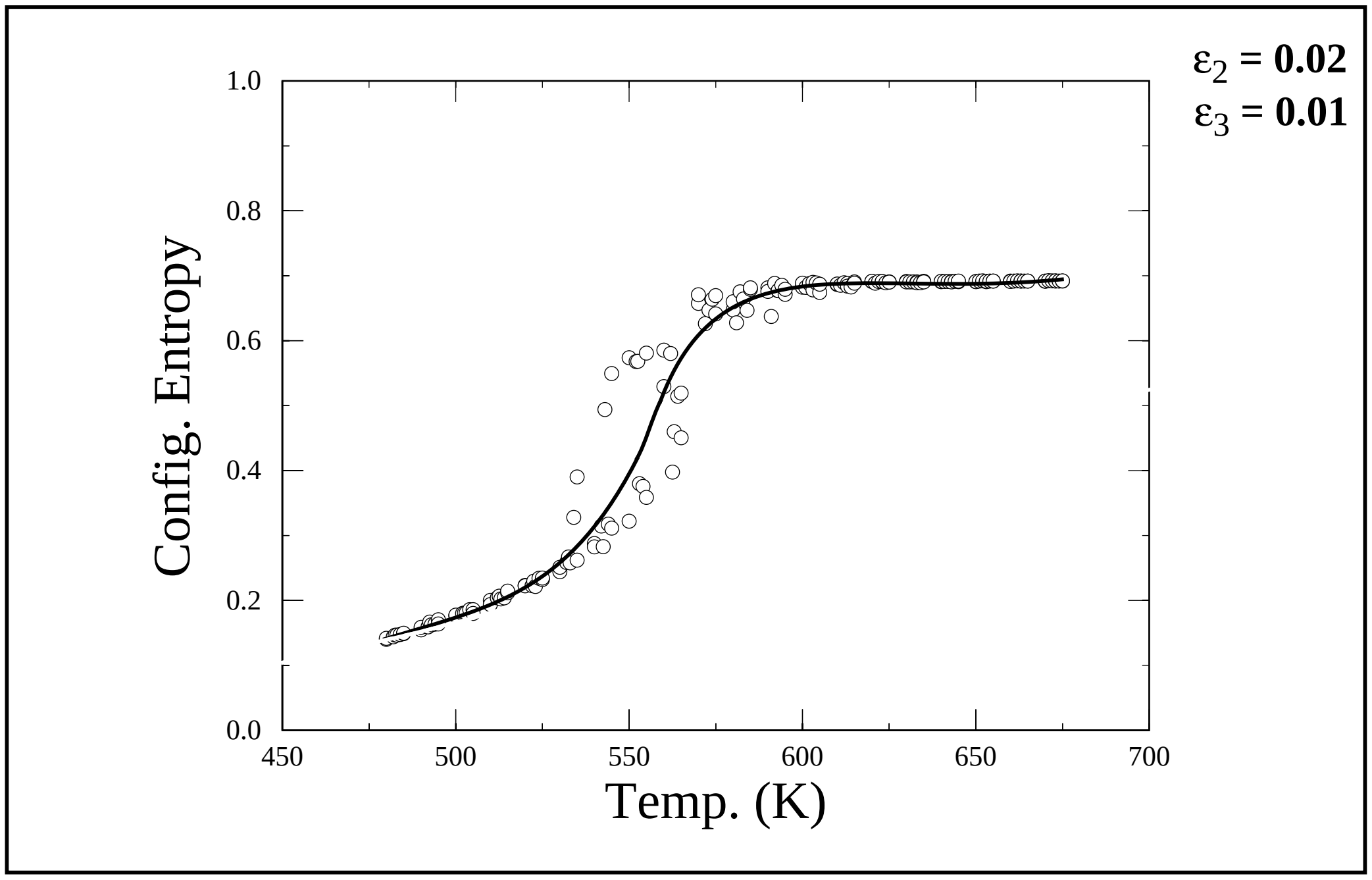

Fig. 4: Configurational entropies from both Onsager's solution and DAM for the system with $\varepsilon_2 = 2\varepsilon_3 = 0.02$, $k = 1\text{x}10^7 k_{B,}$ and $E_{AA} = E_{BB} = E_{AB} = 0$ of Fig. 3.

Let us examine cases in which the order-disorder transition become of first-order kinetics. A system with $\varepsilon_2 = 0.02$, $\varepsilon_3 = 0$, and $E_{AA} = E_{BB} = E_{AB} = 0$ is studied. Vegard's law no longer holds for this system. In Fig. 5(a), specific heat, $C_v$, is presented for both OA (dashed curves) and DAM (solid curves fitted to open circles): the transition temperatures are 968 K for OA and 973 K (= $T_c$) for DAM. In Fig. 5(b), the change in lattice parameter is plotted as a function of temperature: at the order-disorder transition, $\delta$ is shown to increase from about 0.001 to 0.002, clearly indicating a first-order kinetics. This is also in agreement with the prediction of the modified Bragg-Williams approach. Fig. 6 displaces the long range order R in (a) and the short range order $\sigma$ in (b) for the system of Fig. 5. It appears that there is a small drag between the transition in R and the transition temperature, $T_c$, as marked by the vertical line. As noted before, $T_c$ is determined through the sign change in $dC_v/dT$. Any drifting or wobbling of an ordered island should contribute more to the change in R than in $C_v$. The short range order parameter, shown in (b), reveals somewhat a gradual decrease as compared to other thermodynamic properties.

Representing a further deviation from Vegard's law of $\varepsilon_2 = 2\varepsilon_3$, the second system is constituted with $\varepsilon_2 = 0.02$, $\varepsilon_3 = -0.01$, and $E_{AA} = E_{BB} = E_{AB} = 0$. Therefore it is expected to display greater influence from the $R^4$ term in Eq. (3), i.e., a more pronounced first-order kinetics. In Fig. 7, change in lattice parameter (a) and configurational entropy (b) are compared between OA and DAM. Again a small difference is shown in the order-disorder temperature: 2113 K for OA and $T_c$ = 2125 K in DAM. But both methods show excellent agreement in predicting the lattice parameter. DAM's configurational entropies are shown to reach the maximum value of ln2 = 0.693 upon transition on heating, whereas those of OA are still increasing in the disordered state. Clearly, both demonstrate a first-order kinetics, which is further evidenced in the plots of the long range order R and the short range order $\sigma$ in Fig. 8. The DAM data show some hysteretic effects on both heating and cooling cycles.

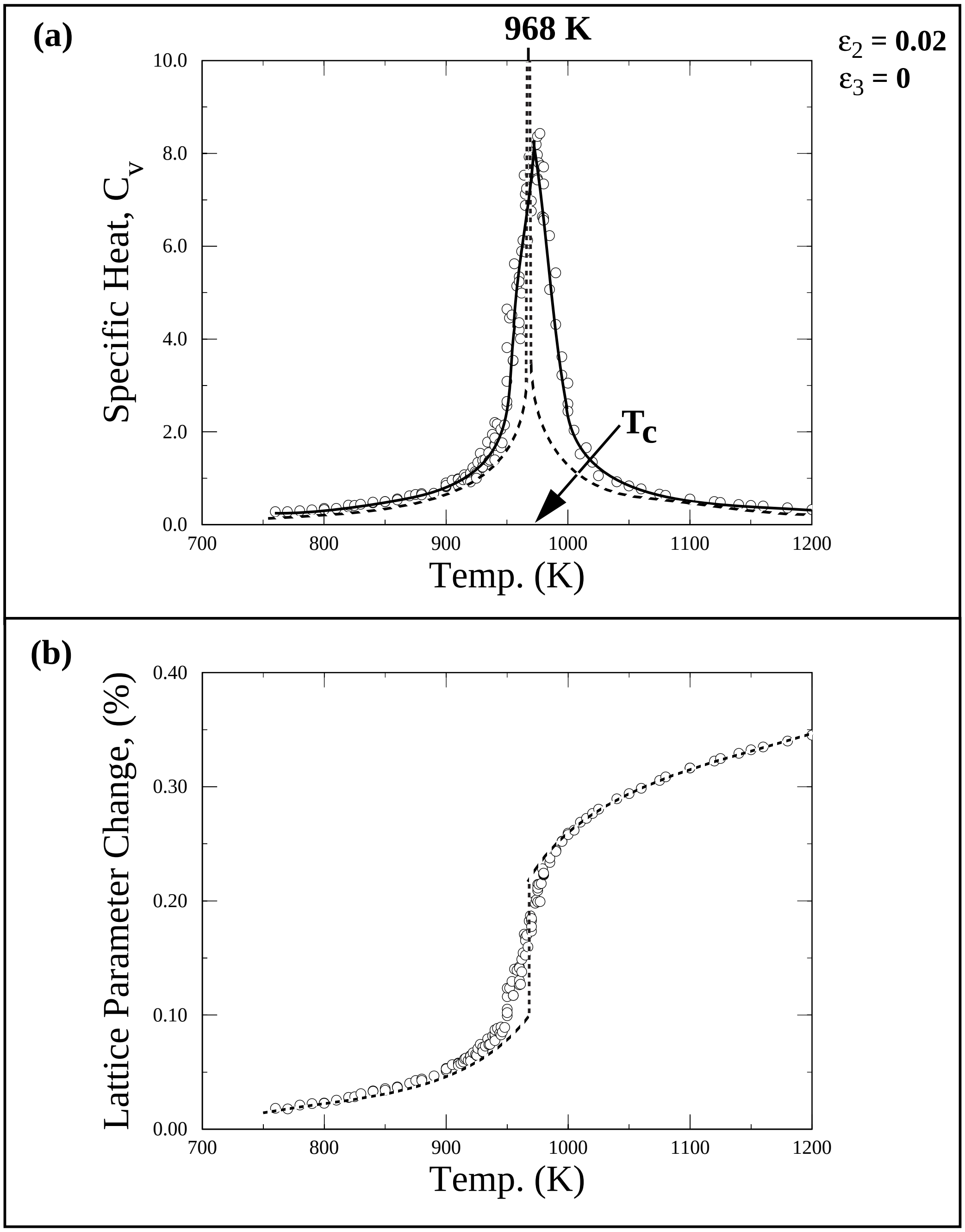

Fig. 5: Order-disorder transition behavior for a two-dimensional square lattice system with $\varepsilon_2$ = 0.02, $\varepsilon_3$ = 0, $k = 1\text{x}10^7 k_B$, and $E_{AA} = E_{BB} = E_{AB} = 0$. In (a), specific heat is plotted as a function of temperature. The dashed curves indicate Onsager's solution, while the solid curve fitted to open circles represents DAM results. DAM's $T_c$ = 973 K. (b) presents a plot of lattice parameter change versus T, clearly displaying a first-order kinetics for the order-disorder transition.

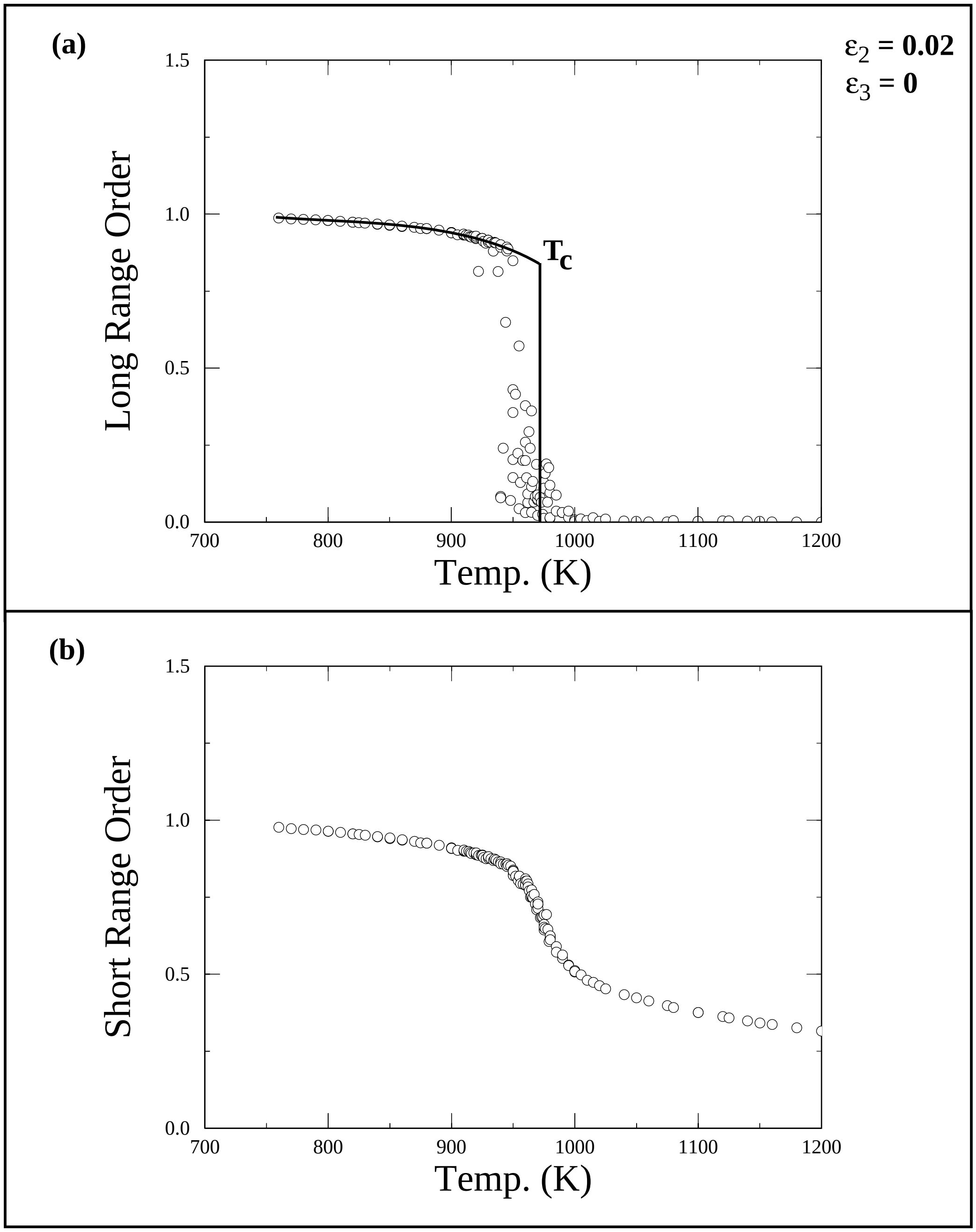

Fig. 6: Long range order R in (a) and short range order σ in (b) for the system with $\varepsilon_2 = 0.02$, $\varepsilon_3 = 0$, $k = 1\text{x}10^7 k_B$, and $E_{AA} = E_{BB} = E_{AB} = 0$ of Fig. 5. All are from DAM, and $T_c$ marks 973 K.

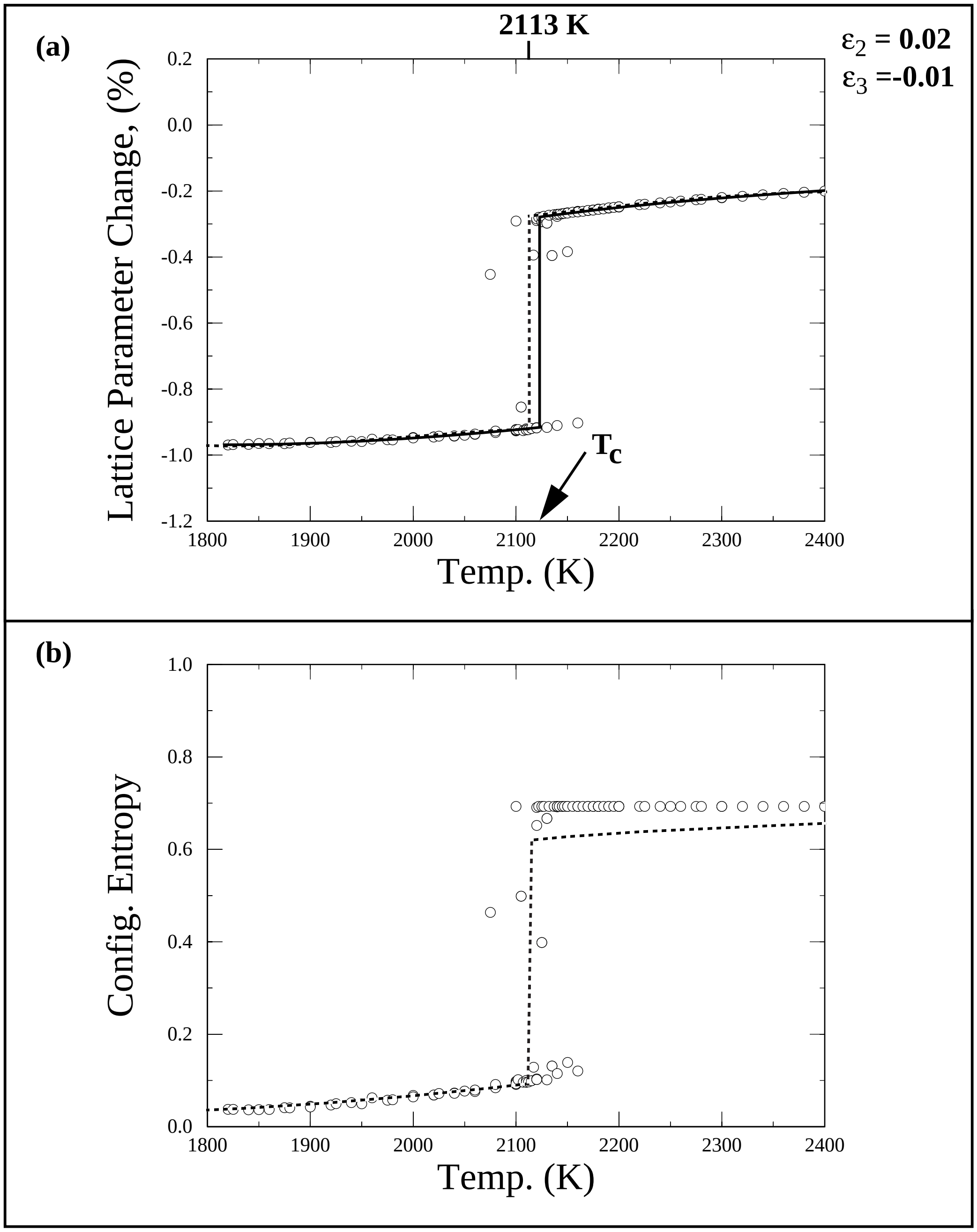

Fig. 7: Order-disorder transition behavior for a system with $\varepsilon_2$ = -2$\varepsilon_3$ = 0.02, $k = 1\text{x}10^7 k_{B,}$ and $E_{AA} = E_{BB} = E_{AB}$= 0. In (a), the change in lattice parameter is plotted as a function of temperature. The dashed curves indicate Onsager's solution, while the solid curve fitted to open circles represents DAM results. $T_c$ marks DAM's transition temperature, 2125 K. (b) compares configurational entropies between Onsager's solution and DAM.

Finally a system with $\varepsilon_2$ = 0.02, $\varepsilon_3$ = -0.01, $E_{AB} = -\Omega_c = -500k_B$, and $E_{AA} = E_{BB} = 0$ is studied: it represents a case with ordering energies both from chemical and from strain origins. Thermal fluctuations are quite large, as shown in Fig. 9(a): OA's $C_v$ shows a much steeper rise at the transition temperature of 2713 K as compared to DAM's $C_v$ with $T_c$ = 2733 K. The variation in lattice parameter indicates a first-order kinetics as shown in Fig. 9(b). The OA transition temperature of 2713 K is higher, by 33 K, than the sum of the individual cases, i.e., 567 K due to $\Omega_c$ and 2113 K due to the strain energy alone. Similarly, the DAM results also show an increase of 33 K in $T_c$. Obviously, there is a coupling effect between the chemical and the strain ordering forces, and this is also expected from the modified Bragg-Williams treatment (see Eq. (3)). Table 1 summarizes the five systems studied: in all, the spring constant is set at $k = 1x10^7 k_B$. All the three cases with first-order kinetics display that the lattice parameter increases from an ordered to a disordered state. The reason is that all the $\varepsilon_3$ are taken to be less than 0.01, the value at the Vegard's law. For example, for the case with $\varepsilon_2$ = 0.02, $\varepsilon_3$ = -0.01, and $\Omega_c = 500k_B$, the lattice parameter increases from 0.991 to 0.996 (see Fig. 9(b)) upon transition on heating. For a system with $\varepsilon_2$ = 0.02, $\varepsilon_3$ = 0.03, and $\Omega_c = 500k_B$, however, the lattice parameter decreases from 1.029 to 1.024 upon transition at the same transition temperature of 2713 K. Therefore, whether an ordered state would have a lesser or a greater lattice parameter than a disordered state depends on the nature of the misfit strains, i.e., the relative values of $\varepsilon_2$ and $\varepsilon_3$.

**Table 1: Comparison between OA and DAM**

| Case | $\Omega_c$ | $\varepsilon_2$ | $\varepsilon_3$ | OA (K) | DAM (K) | Kinetics |
|---|---|---|---|---|---|---|
| 1 | $500k_B$ | 0. | 0. | 567 | 575 | 2nd |
| 2 | 0. | 0.02 | 0.01 | 567 | 580 | 2nd |
| 3 | 0. | 0.02 | 0. | 968 | 973 | 1st |
| 4 | 0. | 0.02 | -0.01 | 2113 | 2125 | 1st |
| 5 | $500k_B$ | 0.02 | -0.01 | 2713 | 2733 | 1st |

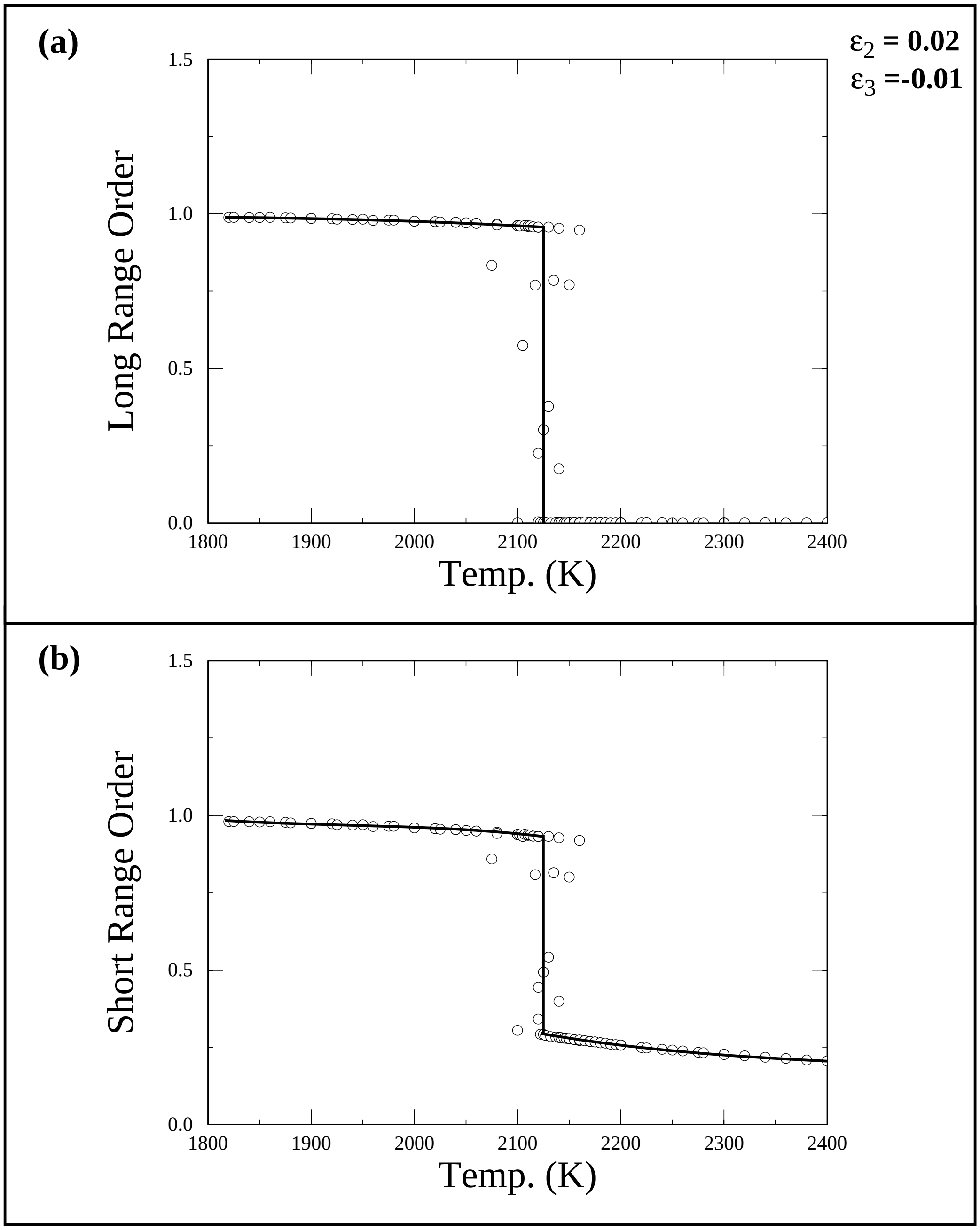

Fig. 8: Long range order R in (a) and short range order σ in (b) for the system with $\varepsilon_2 = 0.02$, $\varepsilon_3 = -0.01$, $k = 1\text{x}10^7 k_B$, and $E_{AA} = E_{BB} = E_{AB} = 0$ of Fig. 7. All are from DAM, and $T_c$ marks 2125 K.

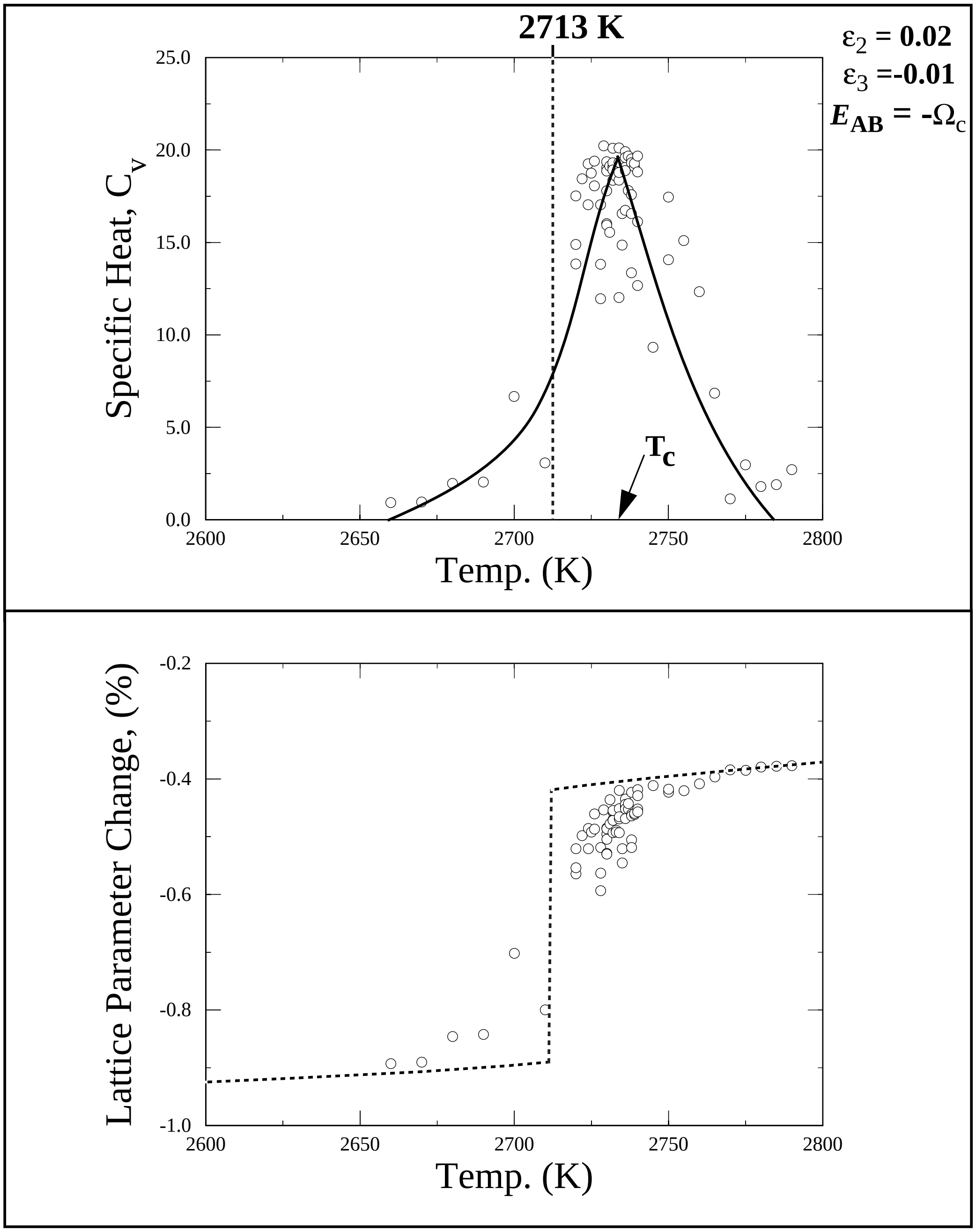

Fig. 9: Order-disorder transition behavior for a system with $\varepsilon_2 = -2\varepsilon_3 = 0.02$, $k = 1\text{x}10^7 k_B$, $E_{AA} = E_{BB} = 0$, and $E_{AB} = -\Omega_c = -500k_B$. In (a), the change in lattice parameter is plotted as a function of temperature. $T_c$ marks DAM's transition temperature, 2733 K. (b) compares change in lattice parameter between Onsager's solution and DAM.

## 6. Discussion and Conclusion

Within the assumption of nearest neighbor interactions on a square lattice, both modified Bragg-Williams and Onsager approach show that coherency strain arising due to atomic mismatch can exert profound effects on order-disorder transitions in substitutional alloys. If the alloy system is elastically homogeneous ($k_1 = k_2 = k_3 = k$) and Vegard's law is obeyed ($\varepsilon_2 = 2\varepsilon_3$), the order-disorder transition is of a second-order kinetics. If the misfit strain $\varepsilon_3$ is significantly different from $0.5\varepsilon_2$, however, the transition may become a first-order kinetics, as the configurational free energy surface is composed of double wells. At the transition of a first-order kinetics, the lattice parameter can either increase or decrease upon heating, i.e., the lattice parameter of an ordered state can be less or greater than that of a disordered state, depending on how $\varepsilon_3$ is deviated from $0.5\varepsilon_2$. The results of Onsager's approach are independently confirmed with those of the Discrete Atom Method.

The present analyses are focused on elastically homogeneous systems. It should be, however, noted that the theories of Bragg-Williams, Onsager, and DAM described above are much more general in treating elastically inhomogeneous systems of any spring constants $k_1$, $k_2$, and $k_3$. In fact, by extending the atomic interactions to second neighbors and thus stabilizing the square lattice, the ground state energy of the ordered structure was recently studied: the results showed that the stability of the ordered structure depends strongly on the elastic anisotropy and Vegard's law represents a limited, special case [19]. In the Ising model, a cross energy term of A-B bond, $E_{AB}$, enters as a thermodynamic identity. In a similar way, both a spring constant $k_3$ and a misfit strain $\varepsilon_3$ are necessary to account for the elastic interaction between A and B species. Obviously, one would question what might be the values of $k_3$ and $\varepsilon_3$ for a given binary system, but the same question can be raised for $E_{AB}$. It is, however, hoped that some first-principles calculations combined with x-ray experiments shed light on these variables in the future. Both original Bragg-Williams and Onsager approaches are not devised for a phase separation. Therefore, the modified versions should be viewed with caution, as they fail in certain aspects: for example, when $\varepsilon_2 = 0$ but $\varepsilon_3 \neq 0$, either Eq. (3) or Eq. (4) fails to predict a phase separation or a clustering behavior, which can be easily detected through the topological evolution in a DAM simulation.

What if local atomic relaxations are allowed? In other words, how useful would the current analyses be? Because of the mechanical stability associated with a square lattice (similarly with a body centered cubic lattice), the atomic interactions must be extended to second neighbors at minimum. Unfortunately, the treatment of second neighbor interactions is a subject beyond both Bragg-Williams and Onsager approach. Currently DAM studies are being performed to answer the question. Some preliminary results with a square lattice indicate that local atomic relaxations are crucial for certain systems such as one with both clustering energy ($E_{AB} > 0$) and strain energy. For other systems where both chemical ordering energy ($E_{AB} < 0$) and strain energy work together, however, local atomic relaxation appears to contribute a minor role in order-disorder transition: one system shows $T_c$ = 1190 K and 1221 K with and without local relaxation, respectively, while another shows $T_c$ = 737 K and 745 K with and without local relaxation, respectively. As one might have expected, local atomic relaxation is shown to lower critical temperatures, but it would be interesting to see to what extent local relaxation would influence ordering behavior in three-dimensional lattices.

**Acknowledgments**

The authors are indebted to Dr. Daniel D. Lee of Bell Laboratories, Lucent Technologies, for enlightening discussions during the course of this work. The work was partially supported with NSF Major Research Instrumentation Grant #9871133.